\documentclass[manuscript, screen, nonacm]{acmart}
\AtBeginDocument{%
  \providecommand\BibTeX{{%
    Bib\TeX}}}

\usepackage{algorithmic}
\usepackage{graphicx}
\usepackage{textcomp}
\usepackage{xcolor}
\usepackage{booktabs}
\usepackage{multirow}
\usepackage{pifont}
\usepackage{tabularx} 
\usepackage{subcaption}
\usepackage{adjustbox}
\usepackage[table]{xcolor}
\colorlet{bestperf}{blue!20}  
\colorlet{bestimprov}{orange!20} 
\usepackage[most]{tcolorbox}

\newtcolorbox{keytakeaway}[1]{
  colback=gray!10,      % Light gray background color
  colframe=black!75,    % Frame color
  fonttitle=\bfseries,
  title={Key Takeaway #1:},
  arc=2mm,              % Rounded corners
  boxrule=1pt,          % Frame thickness
  breakable,            % Allows the box to break across pages
}

\usepackage[ruled, lined, linesnumbered, commentsnumbered, noend]{algorithm2e}
\SetKwComment{Comment}{//}{}
\newcommand{\cmark}{\ding{51}}
\newcommand{\xmark}{\ding{55}}

\def\BibTeX{{\rm B\kern-.05em{\sc i\kern-.025em b}\kern-.08em
    T\kern-.1667em\lower.7ex\hbox{E}\kern-.125emX}}

\begin{document}

\title{DeepV: A Model-Agnostic Retrieval-Augmented Framework for Verilog Code Generation with a High-Quality Knowledge Base}

\author{Zahin Ibnat}
\email{ibnatz16@ufl.edu}
\orcid{0000-0001-5664-4428}
\author{Paul E. Calzada}
\email{paul.calzada@ufl.edu}
\orcid{0000-0002-5001-6321}
\author{Rasin Mohammed Ihtemam}
\orcid{}
\email{r.ihtemam@ufl.edu}
\author{Sujan Kumar Saha}
\orcid{0009-0009-5311-9367}
\email{sujansaha@ufl.edu}
\author{Jingbo Zhou}
\orcid{} % Reminder to fill this in
\email{jingbozhou@ufl.edu}
\author{Farimah Farahmandi}
\orcid{0000-0003-1535-0938}
\email{farimah@ece.ufl.edu}
\author{Mark Tehranipoor}
\orcid{0000-0003-4699-3231}
\email{tehranipoor@ece.ufl.edu}
\affiliation{%
  \institution{University of Florida}
  \city{Gainesville}
  \state{Florida}
  \country{USA}
}

\renewcommand{\shortauthors}{Ibnat et al.}

\begin{abstract}
As large language models (LLMs) continue to be integrated into modern technology, there has been an increased push towards code generation applications, which also naturally extends to hardware design automation. LLM-based solutions for register transfer level (RTL) code generation for intellectual property (IP) designs have grown, especially with fine-tuned LLMs, prompt engineering, and agentic approaches becoming popular in literature. However, a gap has been exposed in these techniques, as they fail to integrate novel IPs into the model's knowledge base, subsequently resulting in poorly generated code. Additionally, as general-purpose LLMs continue to improve, fine-tuned methods on older models will not be able to compete to produce more accurate and efficient designs. Although some retrieval augmented generation (RAG) techniques exist to mitigate challenges presented in fine-tuning approaches, works tend to leverage low-quality codebases, incorporate computationally expensive fine-tuning in the frameworks, or do not use RAG directly in the RTL generation step. In this work, we introduce \emph{DeepV}: a model-agnostic RAG framework to generate RTL designs by enhancing context through a large, high-quality dataset without any RTL-specific training. Our framework benefits the latest commercial LLM, OpenAI's GPT-5, with a near 17\% increase in performance on the VerilogEval benchmark. We host DeepV for use by the community in a Hugging Face (HF) Space: \url{https://huggingface.co/spaces/FICS-LLM/DeepV}.
\end{abstract}

%%
%% Keywords. The author must pick one of the keywords given by the ACM.
\begin{CCSXML}
<ccs2012>
   <concept>
       <concept_id>10010583.10010786.10010787</concept_id>
       <concept_desc>Hardware~Hardware description languages and compilation</concept_desc>
       <concept_significance>500</concept_significance>
       </concept>
   <concept>
       <concept_id>10010147.10010178.10010179</concept_id>
       <concept_desc>Computing methodologies~Natural language processing</concept_desc>
       <concept_significance>500</concept_significance>
       </concept>
 </ccs2012>
\end{CCSXML}

\ccsdesc[500]{Hardware~Hardware description languages and compilation}
\ccsdesc[500]{Computing methodologies~Natural language processing}

\keywords{Verilog, Large Language Models, Retrieval Augmented Generation, RTL Generation}

\maketitle

\section{Introduction} \label{sec:introduction}

Generative artificial intelligence (GenAI) is a cornerstone of modern technology, with its continued usage expanding into various industry sectors as it further enables automation compared to powerful electronic design automation (EDA) tools. Large language models (LLMs), a category of GenAI, have demonstrated impressive results in code generation with faster design output and streamlined workflows~\cite{openai2021codex, liAlphaCode2022}. 

As a result of these successes, hardware engineers have begun to incorporate LLMs into the hardware design flow, especially for the generation of register transfer level (RTL) code for intellectual properties (IPs)~\cite{Abdollahi2025}. However, unlike with software code, LLMs generally struggle to generate high-quality RTL code primarily due to 1) limited high-quality RTL code used in training LLMs, 2) a lack of structural understanding, including timing and parallelism, and 3) a limited understanding of physical design considerations. Many works have attempted to improve LLM-based RTL code generation through prompt engineering~\cite{zhao2025vrankenhancingverilogcode,zhou2023leasttomostpromptingenablescomplex,nair2023generating}, instruction fine-tuning of LLMs~\cite{verigen,chipnemo,rtlcoder,wang2025largelanguagemodelverilog,wu2025itertliterativeframeworkfinetuning,betterv,rtlpp,yubeaton2025verithoughtsenablingautomatedverilog}, or agentic approaches~\cite{AutoChip,tang2024hivegenhierarchicalllmbased,huang2024llmpoweredverilogrtlassistant,gpt4aigchip,autobench,zhang2025llm4dvusinglargelanguage,origen,Coopetitivev,saha2025threatlensllmguidedthreatmodeling,complexvcoder,qin2025reasoningvefficientverilogcode}. However, prompt engineering requires context or examples, despite the rules included, leading to frequent trial-and-error testing~\cite{Geroimenko2025}. Moreover, instruction-tuned LLMs must optimally receive prompts in a specific format, and updating the approach with new data requires retraining, a computationally expensive and time-consuming process~\cite{han2024parameterefficientfinetuninglargemodels}. Similarly, agentic approaches can introduce significant latency and computational cost, as their iterative 'generate-and-verify' cycles require multiple calls to the LLM and external tools for a single output~\cite{Sapkota_2026}.

Recently, a push has been made towards applying retrieval-augmented generation (RAG) approaches with fine-tuned models to enhance LLM-guided hardware design automation tasks. RTLRepoCoder~\cite{rtlrepocoder} utilizes RAG with fine-tuning to enhance repository-level Verilog code completion. AutoVCoder~\cite{autovcoder} leverages both a knowledge and example retriever in addition to a fine-tuned model to generate RTL code. RTLFixer~\cite{rtlfixer} uses RAG and ReAct prompting to interactively debug code with feedback. As models improve and fine-tuning computational costs increase, there is a need for flexible and cost-effective solutions. We are the first to study RAG's effectiveness with a large and high-quality dataset, VerilogDB~\cite{calzada2025verilogdblargesthighestqualitydataset}, without any fine-tuning, showcasing large accuracy gains that meet and rival fine-tuned RTL generator accuracy. Our key contributions are summarized as follows: 

\begin{enumerate}
    \item We introduce \emph{DeepV}: a model-agnostic RTL code generation framework using RAG, where we are the first, to our knowledge, to leverage a dataset of both syntactically correct and synthesizable Verilog codes that has been preprocessed using EDA tools. 
    \item We employ a variety of pre-retrieval optimizations to improve accuracy in addition to implementing dynamic sampling to enhance retriever accuracy and efficiency.
    % \item We show through \emph{DeepV} that complex LLM-based frameworks incur too much overhead and are not necessary for accurate RTL code generation.
    \item We analyze our implementation on VerilogEval~\cite{liu2023verilogevalevaluatinglargelanguage,pinckney2025revisitingverilogevalyearimprovements}, showcasing our approach's capability to improve generation accuracy on diverse design problems.
    \item Additionally, we apply \emph{DeepV} to hierarchical testcases, including FIR filter, Sobel filter, SRNG, and UART, showcasing our tool's capability in generating complex, multimodule IPs. 
\end{enumerate}

Our experiments demonstrate that RAG with VerilogDB~\cite{calzada2025verilogdblargesthighestqualitydataset} enhances the RTL generation capability of the existing models by up to \textit{18.6\%} and \textit{13\%} in the \textit{pass@1} and \textit{pass@10} metric respectively, on the VerilogEval~\cite{liu2023verilogevalevaluatinglargelanguage,pinckney2025revisitingverilogevalyearimprovements} benchmark. Furthermore, \emph{DeepV} outperforms the best-performing state-of-the-art fine-tuned solutions by approximately \textit{10\%}.

\section{Background} \label{sec:background}
\subsection{Automation of Hardware Design: HLS and its Pitfalls} \label{subsec:background_hls}
The challenge of automating hardware design is not new; for decades, the EDA community has attempted to raise the level of abstraction to mitigate the complexities and long development times associated with manual RTL coding. The traditional hardware design flow is a labor-intensive process and represents a significant bottleneck in the development of complex systems-on-chip (SoCs). Before the era of LLMs, the most prominent solution to this challenge was High-Level Synthesis (HLS), which automatically generates RTL code from high-level programming languages like C, C++, or SystemC~\cite{HLS_book}. The goal of HLS is to bridge the gap between design specification and implementation, which increases productivity and allows designers to explore the design space faster within aggressive time-to-market (TTM) constraints~\cite{HLS_book}.

HLS tools promise to further improve productivity by allowing designers to work at a higher level of abstraction, focusing on algorithmic behavior rather than low-level implementation details. However, despite its adoption in certain domains, HLS has faced persistent challenges that have limited its universal application~\cite{Lahti2019}. Designers often struggle to produce RTL that meets the stringent Power, Performance, and Area (PPA) constraints of their designs without extensive manual code restructuring and pragma insertion~\cite{Lahti2019}. Additionally, the abstraction offered by HLS introduces a new verification challenge: ensuring that the automatically generated RTL is functionally equivalent to the original high-level specification. Moreover, the HLS compilation and optimization process can inadvertently introduce security vulnerabilities, such as information leakage or insecure arbitration, which are not present in the source C/C++ code~\cite{Pundir2022, Muttaki2022}. These vulnerabilities often arise during HLS stages like scheduling and resource binding, where the insecure sharing of hardware resources can create side channels or other exploits that compromise security~\cite{Ibnat2023DFT, Shang2023}. This issue has led to extensive research focused on the security assessment of HLS tools and the development of security-aware synthesis flows intended to mitigate these risks~\cite{Muttaki2021, Pundir2021, Shang2023}.

Beyond these specific concerns, the fundamental limitations of HLS, namely the difficulty in achieving optimal PPA without extensive manual intervention and the rigid requirement for specifications to be written in a formal programming language, have persisted. These challenges highlighted the need for a new paradigm in design automation capable of interpreting high-level intent directly from natural language and using the vast knowledge encoded in existing, human-written RTL designs. The emergence of LLMs, with their powerful capabilities in natural language understanding and pattern recognition in code, provided a promising new direction to address these long-standing goals.

\subsection{Foundational Models for Code Generation}\label{subsec:background_foundational_models}

The recent advancements in LLM-based RTL generation are built upon a foundation of powerful, general-purpose code intelligence models. These models are based on the Transformer architecture and are pre-trained on vast corpora of publicly available source code from diverse programming languages~\cite{feng2020codebertpretrainedmodelprogramming}. This large-scale pre-training gives them a fundamental understanding of programming syntax, structure, and semantics. Pioneering efforts, such as OpenAI's Codex, the model underlying GitHub Copilot and Meta's open-source Code Llama, showed that LLMs could achieve great performance on a wide range of code-related tasks, including completion, translation, and generation from natural language prompts~\cite{openai2021codex, roziere2024code}.

These foundational models provide the basis for the two primary methodologies used to create specialized solutions for Verilog generation:

\begin{itemize}
    \item \textbf{Prompt Engineering}: Prompt engineering involves carefully designing the input prompt given to a pre-trained LLM to guide its output toward the desired result~\cite{sahoo2025systematicsurveypromptengineering}. It treats the LLM as a black box and requires no changes to the model's underlying weights. The quality of the generated code is highly dependent on how well the prompt is structured for the intended task.
    \item \textbf{Fine-Tuning}: Fine-tuning adapts a pre-trained model to a specific domain by continuing the training process on a smaller, curated dataset~\cite{ANISUZZAMAN2025100184}. For RTL generation, this involves fine-tuning a base code model on a high-quality corpus of Verilog. This process modifies the model's weights, embedding domain-specific knowledge directly into the model to improve its accuracy and fluency in the target language. While effective, updating all of a model's billions of parameters during full fine-tuning is computationally intense. To address this, Parameter-Efficient Fine-Tuning (PEFT) methods have become standard practice, especially within the domain of fine-tuning for code generation~\cite{han2024parameterefficientfinetuninglargemodels}.
\end{itemize}

\subsection{LLM for RTL Code Generation}
\label{subsec:background_rtl_gen} 
The advancements in LLMs have resulted in a significant shift in the EDA sector, providing better automation of complex hardware design tasks to combat aggressive TTM constraints~\cite{ALSAQER2024}. By using natural language descriptions to generate RTL code, LLMs can shorten design cycles, lower the barrier to entry for hardware development, and enhance designer productivity~\cite{Saha2024}. The exploration of LLMs for hardware has rapidly evolved throughout the past few years, moving from initial prompting experiments to the development of highly specialized, domain-specific models and agentic frameworks.

Early research primarily focused on evaluating general-purpose LLMs through prompt engineering. Researchers investigated how to formulate natural language specifications to achieve more accurate and functionally correct Verilog outputs from models not explicitly trained for hardware design~\cite{chip-chat, chipgpt}. These works enabled more complex, automated frameworks that used LLMs as reasoning engines within a larger design loop to automate design processes~\cite{AutoChip, gpt4aigchip}. While it was proven that prompt engineering can make an impact on the generated RTL code, the early works also showcased the limitations of relying solely on prompting: the process often required extensive trial-and-error, and the models' lack of domain-specific knowledge resulted in frequent syntactic errors.

To overcome these limitations, the focus was shifted toward domain-specific fine-tuning, enabled by the creation of high-quality Verilog datasets. Notable examples of these datasets include RTLLM~\cite{lu2024rtllm} and OpenLLM-RTL~\cite{liu2024openllm}, which provide benchmarks and datasets for LLM-aided RTL generation; MG-Verilog~\cite{zhang2024mgverilogmultigraineddatasetenhanced}, which introduced a multi-grained dataset to enhance generation capabilities; CraftRTL~\cite{craftrtl}, which focused on generating high-quality synthetic data; and VerilogDB, the largest and highest-quality dataset for training and evaluation~\cite{calzada2025verilogdblargesthighestqualitydataset}.

As discussed, these datasets allowed for a new wave of LLM-based solutions for RTL code with specialized, fine-tuned models. Verigen~\cite{verigen} was an early example, fine-tuned on a curated dataset of Verilog from GitHub repositories and academic materials. RTLCoder~\cite{rtlcoder} presented itself as a lightweight, open-source solution that utilized a specialized dataset to outperform larger, general-purpose models like GPT-3.5 on RTL generation tasks. Further advancements in fine-tuning have incorporated more advanced techniques. For example, BetterV~\cite{betterv} utilizes discriminative guidance and reward models tied to hardware performance metrics to control the generation process, while RTL++~\cite{rtlpp} enhances a model's semantic understanding by incorporating control and dataflow graph embeddings into the training process.

Most recently, the field has advanced toward complex, agentic frameworks that contain a multi-step design and verification workflows. These systems are far more advanced than single-shot code generation; instead, they create iterative loops of generation, feedback, and refinement. AutoChip~\cite{AutoChip} was an early framework that automated HDL generation using iterative feedback from the LLM. OriGen~\cite{origen} uses a self-reflection mechanism with a dual-model structure, where one model generates code and another corrects it based on compiler feedback. Other systems integrate with standard EDA tools to create verification loops. For example, some frameworks focus on self-verification and self-correction by simulating or synthesizing the generated code and feeding the results back to the LLM for debugging~\cite{huang2024llmpoweredverilogrtlassistant}. CoopetitiveV~\cite{Coopetitivev} has a multi-agent prompting strategy where different LLM agents collaborate to improve code quality. The development of these advanced systems is supported by a simultaneous effort in benchmarking. Frameworks like VerilogEval~\cite{liu2023verilogevalevaluatinglargelanguage, pinckney2025revisitingverilogevalyearimprovements} and its successors~\cite{pinckney2025comprehensiveverilogdesignproblems} provide standardized problems to measure progress across different models and methodologies, ensuring that the field can systematically evaluate and compare new approaches. 

\subsection{Retrieval Augmented Generation (RAG)} \label{subsec:RAG}

RAG is a powerful technique for enhancing the accuracy and reliability of LLMs by grounding their responses in external, verifiable knowledge~\cite{gao2024retrievalaugmentedgenerationlargelanguage,koziolek_rag}. Instead of relying solely on the parametric knowledge learned during training, RAG dynamically augments relevant, real-time information into the user prompt, providing the LLM with applicable context to make its output. For RTL code generation, this external knowledge consists of high-quality, trusted Verilog designs and associated metadata, which serve as functional examples to guide the generation process~\cite{autovcoder}.

The RAG pipeline works in two phases. First, in an offline indexing stage, a knowledge base of relevant documents, such as verified IP cores, design specifications, or code snippets from repositories like OpenCores~\cite{opencores} and GitHub is prepared. Each document is chunked and converted into a numerical vector representation by an embedding model. These vectors are then stored in a specialized vector database, optimized for efficient similarity search using algorithms like FAISS \cite{FAISS}. Second, during the online generation phase, a user's query is also converted into a vector. This query vector is used to search the database and retrieve the most relevant document chunks (e.g., the top-k most similar code examples). These retrieved examples are then augmented with the original user prompt and fed to the LLM, which uses this new context to generate the final RTL output.

Compared to supervised fine-tuning, RAG offers several advantages for domain-specific tasks like RTL generation:
\begin{itemize}
    \item \textbf{Cost-Effectiveness and Adaptability}: Fine-tuning is a computationally expensive process that requires retraining the entire model to incorporate new data~\cite{ANISUZZAMAN2025100184}. In contrast, a RAG system's knowledge base can be easily updated by simply adding new documents to the vector database, allowing the system to adapt without modifying the LLM's weights.
    \item \textbf{Reduced Hallucination}: By grounding the LLM in a corpus of factual, high-quality code examples, RAG mitigates the risk of hallucination, where the model might generate syntactically correct but functionally incorrect or nonsensical Verilog.
    \item \textbf{Traceability and Interpretability}: When a fine-tuned model generates code, its reasoning is generally unknown. A RAG system, however, can provide references to the specific documents it retrieved to generate a response. This traceability is particularly useful in high-assurance applications, as it allows engineers to verify the source of the retrieved code.
    \item \textbf{Dynamic Adaptation}: RAG provides context that is tailored to each specific prompt. This allows for better generalization across a wide variety of design problems, whereas a fine-tuned model's knowledge is static and fixed at the time of training.
\end{itemize}

The efficiency of a RAG system for code generation is contingent on three factors: 1) the quality and comprehensiveness of the code corpus, 2) the precision and recall of the document retriever, and 3) the reasoning capability of the generator LLM. While some works have successfully combined RAG with fine-tuning~\cite{rtlrepocoder}, the key question we seek to answer is whether a RAG system with a very high-quality code corpus and a highly precise retriever can generate RTL code as accurately, or more accurately, than state-of-the-art fine-tuned models.

\begin{figure}[htbp]
    \centering
    \includegraphics[width=\textwidth]{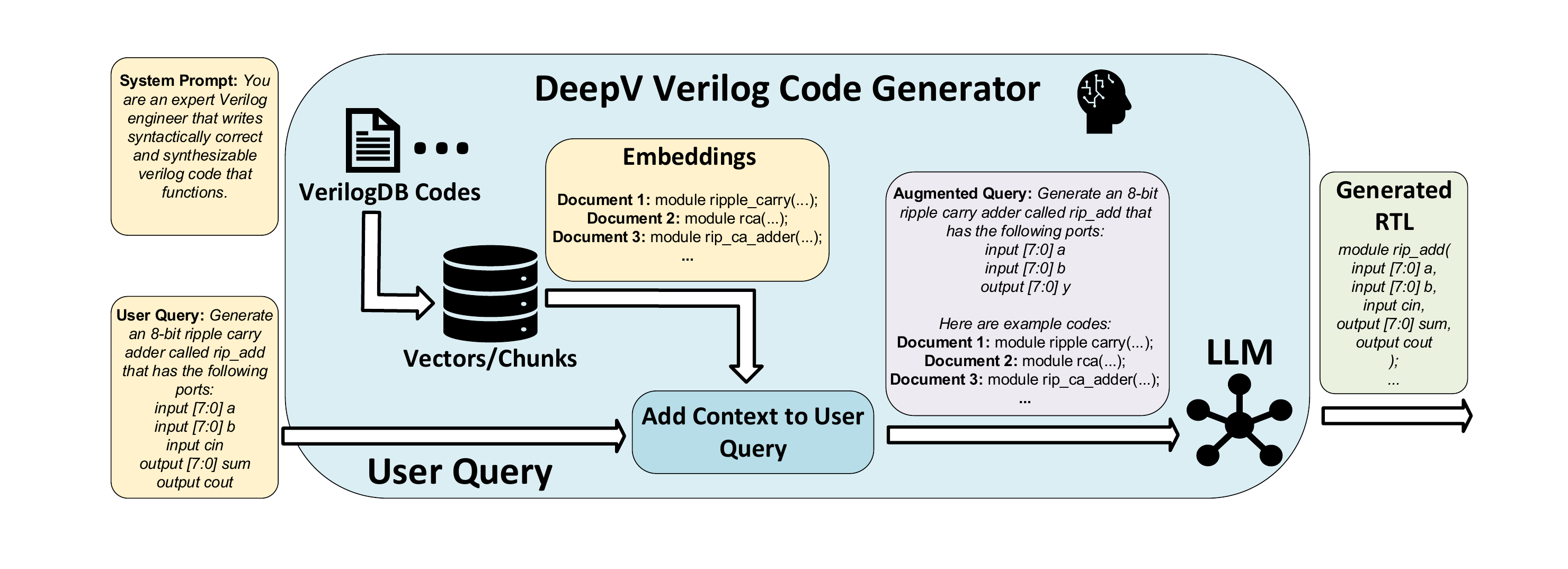}
    \caption{\emph{DeepV} Flow highlighting retrieval augmented generation of an example ripple carry adder design.}
    \label{fig:DeepRAG_framework}
\end{figure}

\section{DeepV} \label{sec:methodology}
To address the challenges of generating accurate Verilog, we introduce \emph{DeepV}, a model-agnostic RAG framework designed to enhance the capabilities of LLMs for RTL design. General-purpose LLMs often struggle with RTL design, frequently hallucinating code that is syntactically invalid or not synthesizable into functional hardware, a limitation the RAG approach is designed to mitigate as discussed in Section~\ref{subsec:RAG}. The primary function of \emph{DeepV} is to aid the chosen LLM with a high-quality knowledge base of existing Verilog designs, providing relevant, in-context examples to guide the code generation process for any given user query. Because \emph{DeepV} is model-agnostic, it does not rely on internal weights or specific API of any single LLM, only requiring the standard API for querying a given LLM with a sufficient context window size. As illustrated in Fig.~\ref{fig:DeepRAG_framework}, the \emph{DeepV} pipeline is composed of three stages: 1) Knowledge Base Preparation, 2) Dynamic Context Retrieval, and 3) Augmented Code Generation. First, a comprehensive Verilog example corpus is prepared and vectorized. Second, these vectorized code snippets are stored in an FAISS index to enable dynamic retrieval of the most relevant examples for a user's prompt. Finally, the retrieved examples are combined with the original query to create an augmented prompt, which is then passed to the LLM to generate the final RTL code. The methodology allows for a given LLM to be consistently provided high-quality, validated examples, allowing in-context lerarning to generate syntactically correct and functionally verifiable RTL codes.

A comparison of our work and notable academic solutions can be seen in Table~\ref{tab:soa_methods}. Namely, our \emph{DeepV} follows a RAG-based approach without complex fine-tuning, opposite to that of the AutoVCoder~\cite{autovcoder}, ComplexVCoder~\cite{complexvcoder}, and RTLRepoCoder~\cite{rtlrepocoder}. Moreover, unlike every other work in the table, \emph{DeepV} is model-agnostic, meaning that it does not depend on the model parameters of any singular LLM. Lastly, \emph{DeepV}'s knowledge base undergoes post-processing with syntax and synthesis checks in addition to metadata extraction and refinement by OpenAI's GPT. To showcase these major points, the following subsections will provide a detailed description of each component of this framework.

\begin{table}[htbp]
\centering
\caption{Comparison of RTL Code Generation Methods}
\label{tab:soa_methods}
\renewcommand{\arraystretch}{1.25} 
\small

\begin{tabular}{
    >{\raggedright\arraybackslash}p{0.18\linewidth} % Method
    >{\raggedright\arraybackslash}p{0.35\linewidth} % Dataset (widest column)
    c % RAG
    c % Fine-Tuned
    >{\raggedright\arraybackslash}p{0.25\linewidth} % Base Model(s)
}
\toprule
\textbf{Method} & \textbf{Dataset} & \textbf{RAG} & \textbf{Fine-Tuned} & \textbf{Base Model(s)} \\
\midrule
\multicolumn{5}{l}{\textit{\textbf{Fine-Tuning Dominant Approaches}}} \\
\midrule
Verigen~\cite{verigen} & 
    \textbf{Source}: GitHub, textbook \newline 
    \textbf{Process}: Filtered 
    & \xmark & \cmark & CodeGen-16B \\
RTLCoder~\cite{rtlcoder} & 
    \textbf{Size}: 27k; \textbf{Source}: Synthetic \newline 
    \textbf{Process}: Syntax checked 
    & \xmark & \cmark & Mistral-7B, DeepSeek-6.7B \\
Origen~\cite{origen} & 
    \textbf{Size}: 220k; \textbf{Source}: Synthetic \newline 
    \textbf{Process}: Syntax checked 
    & \xmark & \cmark & DeepSeek-7B \\
CodeV~\cite{codev} & 
    \textbf{Size}: 165k; \textbf{Source}: GitHub \newline 
    \textbf{Process}: Syntax checked 
    & \xmark & \cmark & CodeLlama-7B+, Qwen1.5-7B \\
VeriCoder & 
    \textbf{Size}: 125k+; \textbf{Source}: Synthetic \newline 
    \textbf{Process}: Functionally validated 
    & \xmark & \cmark & Qwen, DeepSeek \\
CraftRTL~\cite{craftrtl} & 
    \textbf{Size}: 86k+; \textbf{Source}: Synthetic, GitHub \newline 
    \textbf{Process}: Syntax checked 
    & \xmark & \cmark & CodeLlama-7B, Starcoder2-15B \\
BetterV~\cite{betterv} & 
    \textbf{Size}: 68k; \textbf{Source}: GitHub \newline 
    \textbf{Process}: Syntax checked 
    & \xmark & \cmark & CodeLlama-7B, DeepSeek-6.7B, Qwen1.5-7B \\
RTL++~\cite{rtlpp} & 
    \textbf{Size}: 200k+; \textbf{Source}: GitHub+ \newline 
    \textbf{Process}: GPT-refined, synthesis checked 
    & \xmark & \cmark & CodeLlama-7B \\
\midrule
\multicolumn{5}{l}{\textit{\textbf{RAG and Hybrid Approaches}}} \\
\midrule
AutoVCoder~\cite{autovcoder} & 
    \textbf{Source}: Synthetic, GitHub \newline 
    \textbf{Process}: Scoring filtered 
    & \cmark & \cmark & CodeLlama-7B, DeepSeek-6.7B, Qwen1.5-7B \\
ComplexVCoder~\cite{complexvcoder} & 
    \textbf{Size}: 12.5k RAG base \newline 
    \textbf{Source}: GitHub, Pyranet~\cite{pyranet}
    & \cmark & \cmark & DeepSeek-V3, GPT-4o, Qwen2.5 \\
RTLRepoCoder~\cite{rtlrepocoder} &
    \textbf{Source}: Repository-level context \newline
    \textbf{Process}: Cross-file parsing
    & \cmark & \cmark & Code Llama, StarCoder \\
\midrule
\multicolumn{5}{l}{\textit{\textbf{Agentic and Feedback-Driven Approaches (No Fine-Tuning)}}} \\
\midrule
AutoChip~\cite{AutoChip} & 
    \textbf{Source}: N/A (Feedback-based)
    & \xmark & \xmark & GPT (code-davinci-002) \\
RTL-Assistant~\cite{huang2024llmpoweredverilogrtlassistant} &
    \textbf{Source}: N/A (Feedback-based)
    & \xmark & \xmark & GPT-4, GPT-3.5 \\
VeriThoughts~\cite{yubeaton2025verithoughtsenablingautomatedverilog} &
    \textbf{Source}: N/A (Feedback-based)
    & \xmark & \xmark & GPT-4 \\
VeriMind &
    \textbf{Source}: N/A (Agentic reasoning)
    & \xmark & \xmark & GPT-4 (or similar) \\
CoopetitiveV~\cite{Coopetitivev} &
    \textbf{Source}: N/A (Prompting strategy)
    & \xmark & \xmark & GPT-4, Claude 3, Gemini 1.5 \\
\midrule
\textbf{DeepV} & 
    \textbf{Size}: 20k+; \textbf{Source}: GitHub, OpenCores, textbook \newline \textbf{Process}: GPT-refined, Syntax + synthesis checked 
    & \cmark & \xmark & \textbf{Model Agnostic} \\
\bottomrule
\end{tabular}
\end{table}

\subsection{Verilog Example Corpus and Vectorization}\label{subsec:corpus}

The performance of a RAG system is fully dependent on the quality and comprehensiveness of its knowledge base. VerilogDB~\cite{calzada2025verilogdblargesthighestqualitydataset} was selected as the Verilog example corpus. VerilogDB provides the largest, high-quality dataset specifically curated for LLM-based RTL generation. It is composed of 20,392 Verilog modules (totaling 751 MB of code) sourced from repositories like GitHub and OpenCores, as well as academic materials, which totaled to be over 30GB of raw Verilog data. The collection of modules has a broad coverage across main RTL design classes, including combinational and sequential logic, custom accelerators, peripheral interfaces, and basic digital building blocks. Additionally, every module in the database has undergone a rigorous preprocessing framework to verify that it is both syntactically correct and synthesizable with standard EDA tools, ensuring the examples reflect industry-ready design practices. Specifically, the process includes syntax validation using the \emph{Icarus Verilog (iVerilog)} compiler and logic synthesis checking via \emph{Yosys}, making sure that the database is free of common HDL errors and unsynthesizable constructs often resulting from the accidental inclusion of SystemVerilog constructs in Verilog-2005 code. Furthermore, each module is accompanied by a natural language description, which is necessary for an accurate similarity search against a user's query.

\begin{figure}[htbp]
    \centering
    \includegraphics[width=\columnwidth]{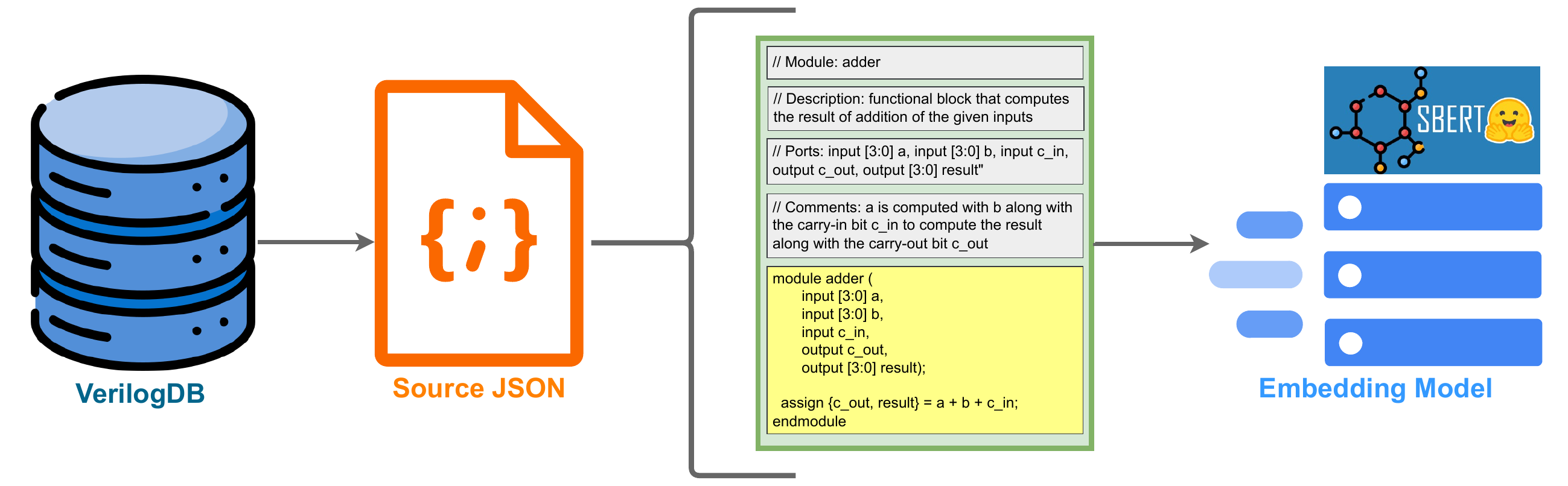}
    \caption{Document Construction for Embedding}
\label{fig:document_construction}
\vspace{0.1in}
\end{figure}

To create a well-detailed representation for each design, a single document is constructed for every module in the database, as seen in Fig.~\ref{fig:document_construction}. This document is not just the raw code; it is a structured text block that begins with header information (including the module's name, its natural language description, a detailed list of its ports, and all original code comments). Structuring the document with this descriptive metadata and natural language first allows for the maximizing of the semantic relevance captured by the embedding model's attention window and improves query-to-document correlation by aligning with the natural language format of the user's query. This header, formatted as Verilog comments, is followed by the full Verilog code of the module within the dedicated JSON for each document. This method makes it so that each document contains the complete context, both structural and descriptive, of a single hardware module. Adding this metadata required little overhead as the corpus is primarily composed of small to moderate designs where the majority of modules contained fewer than 100 lines of code, which is ideal for the RAG task as the examples are self-contained, high-quality, and computationally manageable for in-context learning.

To handle the scale of the entire 20,392-module corpus, the documents are processed in sequential, memory-efficient batches. Each structured document within a batch is then converted into a high-dimensional vector using the \texttt{all-MiniLM-L6-v2} embedding model. This is a highly efficient model from the Sentence-Transformers framework, which uses knowledge distillation to create a small (6-layer) but powerful model optimized for generating semantically meaningful document embeddings~\cite{reimers-2019-sentence-bert}. The \texttt{all-MiniLM-L6-v2} model outputs a 384-dimensional vector. This size is a balance between retrieval accuracy and memory footprint, allowing for low-latency inference and minimizing vectorization overhead, which prevents a latency bottleneck for the real-time RAG query. The final output of this stage is a complete vector representation of the VerilogDB corpus, where each vector captures the rich semantic meaning of an individual hardware module, making it ready for indexing and retrieval.

\subsection{FAISS Storage and Dynamic Retrieval}\label{subsec:FAISS}
Once the database is vectorized, the resulting high-dimensional vectors are stored in a Facebook AI Similarity Search (FAISS) index. FAISS is an open-source library specifically designed for rapid similarity search in massive, high-dimensional datasets~\cite{douze2025faisslibrary}. It allows for the comparison of a user's query against all 20,392 document vectors in real-time. At initialization, the system loads this pre-built FAISS index into memory, preparing the retrieval for incoming user queries. Given the current corpus size of 20,392 modules, an exact search approach using the \texttt{IndexFlatL2} structure was selected via the LangChain FAISS wrapper, guaranteeing 100\% retrieval accuracy without the speed-vs-recall trade-offs associated with approximate nearest neighbor indices like InVerted File Flat (IVFFlat) or hierarchical navigable small worlds (HNSW).

The retrieval of relevant examples is not a simple top-k lookup; it is a multi-stage dynamic retrieval process that finds the most contextually valuable documents for prompt augmentation that is unique to each query. In short, this process acts as a sophisticated filter. It is initiated when the user's natural language query is converted into a high-dimensional vector using the same embedding model that was used for the corpus. This query vector is then used to perform a similarity search against the FAISS index.

The system calculates the Euclidean Distance (L2) between the query vector and each document vector, which measures the straight-line distance between them in the embedding space; a lower distance indicates a closer match~\cite{euclidean2004}. While Cosine Similarity is commonly used for directional alignment in vector spaces, L2 distance was chosen as the underlying measure, which is mathematically equivalent to Cosine Similarity for normalized vectors, while providing a direct, un-normalized distance metric for subsequent thresholding purposes. The LangChain framework then converts these distance measurements into a normalized relevance score, where a higher value signifies greater similarity, to provide a consistent metric for the subsequent filtering stages~\cite{langchain2023}. This search retrieves an initial candidate pool of the 10 documents with the highest converted relevance scores. Following this initial retrieval, the system applies a filter on the candidate pool based on an absolute relevance score. A minimum score threshold is enforced, and any document with a relevance score below 0.55 is discarded from the pool, ensuring that only documents with a strong similarity to the user's request are considered for the final context.

The final stage of the retrieval process is a dynamic sampling method designed to adapt the number of examples to the specific nature of the user's query. This process can be seen in Alg.~\ref{alg:dynamic_sampling}, where instead of using a fixed number of documents, the system analyzes the rate of decrease in relevance scores among the sorted candidates. It identifies the point of diminishing returns by calculating the drop in score between each consecutively ranked document. If the score drop between two documents is more than 1.5 times larger than the drop between the previous two, the system halts the selection process. The specific values of the 0.55 relevance threshold and the 1.5× dynamic drop factor were tuned on a validation set to optimize for both context quality and LLM token budget. This heuristic allows the context to be flexible; for queries with many highly relevant examples, more documents may be included, while for queries where relevance drops off sharply, the context will be more focused. As a final safeguard to manage the context window size and associated computational costs, the number of documents selected by this dynamic process is ultimately capped at a maximum of five examples. This multi-stage retrieval strategy ensures that the context provided to the LLM is not only highly relevant but also dynamically sized to best match the complexity of each user query.

The usage of this algorithm as well as specific val for each variable in Alg.~\ref{alg:dynamic_sampling} can be seen in Section~\ref{sec:evaluation}. As a point of discussion, it can be stated that the dynamic sampling algorithm is dependent upon the FAISS system's scoring, which is as stated, the Euclidean distance between the query vector and each document vector. While this is a consistent metric and a simple scoring method, it can prove to be unreliable in some cases. Not all prompts will be set up in a way that matches the document's metadata for the description of the IP. While the default scoring system from FAISS has proven to be reliable in majority of cases, other scoring metrics can be evaluated in subsequent works for performance gains.  In particular, future work could explore implementing the IVFFlat to scale to larger corpora beyond 100,000 examples, accepting a minor trade-off in retrieval latency for massive dataset management.

\newcommand{\Var}{\textup}
\newcommand{\algorithmicbreak}{\textbf{break}}
\newcommand{\BREAK}{\STATE \algorithmicbreak}
\SetKwInput{KwRequire}{Require}
% Set algorithm line numbers
\SetNlSty{}{}{:}
\setlength{\textfloatsep}{0pt}% Remove \textfloatsep

\begin{algorithm}[htbp]
    \KwRequire{
    \\ $D = \{(d_1, s_1), \dots, (d_n, s_n)\}$: List of candidates sorted by relevance score ($s_1 \ge s_2 \ge \dots$).\\
    $\tau$: The minimum relevance score threshold.\\
    $k_{max}$: The maximum number of documents to select (e.g., 5).\\
    $\alpha$: The score drop-off factor (e.g., 1.5).}
    \KwResult{$C$: The final list of context documents.}
    \Comment{1. Filter documents below the minimum threshold}
    $D' \leftarrow \{(d, s) \in D \mid s \ge \tau\}$\; 
    \Comment{2. Initialize empty context}
    $C \leftarrow \emptyset$\; 
    \If{$|D'| \ge 1$}{%
        \Comment{Always include the most relevant document}
        $C \leftarrow C \cup \{d'_1\}$ 
    }
    \If{$|D'| \ge 2 \text{ and } k_{max} \ge 2$}{%
        \Comment{Include the second document to establish the initial drop}
        $C \leftarrow C \cup \{d'_2\}$\; 
        \Comment{Calculate the first score drop}
        $\Delta_{prev} \leftarrow s'_1 - s'_2$\; 
    
        \For{$i \gets 3 \text{ to } \min(|D'|, k_{max})$}{%
            \Comment{Calculate the current score drop}
            $\Delta_{curr} \leftarrow s'_{i-1} - s'_i$\; 
            \If{$\Delta_{curr} > \alpha \cdot \Delta_{prev}$}{% 
                \Comment{Halt if relevance drops off sharply}
                \textbf{break}\;
            }
        \Comment{Add the current document to the context}
        $C \leftarrow C \cup \{d'_i\}$\;
        \Comment{Update previous drop for the next iteration}
        $\Delta_{prev} \leftarrow \Delta_{curr}$\; 
       }
    }
    \KwRet $C$
    \caption{Dynamic Sampling for RAG Context}
    \label{alg:dynamic_sampling}
\end{algorithm}

\subsection{RTL Code Generation} \label{subsec:RTL_code_generation}
With the FAISS index constructed with an algorithm for a dynamic sampling RAG, the final stage of the \emph{DeepV} framework is the generation of RTL code. The backend for the framework is an LLM of the user's choosing, as our methodology is model-agnostic. Additionally, the dynamic retriever works alongside meticulous prompt engineering that is designed to constrain the LLM's output to adhere to strict hardware design standards. In the case of a commercial model such as GPT-4.1, the system prompt first assigns a role to the model to act as an expert Verilog engineer who adheres to strict rules for the generation task. These include, but are not limited to:
\begin{itemize}
    \item You must produce fully implemented, accurate Verilog-2005 code.
    \item You must not use placeholders or incomplete modules.
    \item You cannot nest a module inside another module.
\end{itemize}

Following the system prompt, the user prompt combines the retrieved documents with the original query. The complete augmented prompt is constructed as a three-part sequence: 1) the System Role and Constraints, 2) the Contextual Code Examples (retrieved by RAG), and 3) the User's Natural Language Request. The Verilog modules selected by the dynamic retrieval process are augmented into the query as contextual examples, which can add to expected coding style and structure and the intended function of the requested module. The complete prompt applies in-context learning, where the LLM uses the provided high-quality documents to inform its generation of the new module. Then, the structured prompt is sent to the LLM via an API endpoint. After the model generates a response, a final post-processing step is applied to ensure the output is clean and directly usable. This step uses regular expressions to parse the raw text and extract only the Verilog code from the expected markdown block, isolating it from any extraneous text the model might produce. This extracted code is the final, ready-to-use RTL design produced by the framework.

\section{Evaluation} \label{sec:evaluation}
To cover a wide range of designs as well as provide quantitative analysis for \emph{DeepV}'s performance, the evaluation for the framework utilized the VerilogEval benchmark~\cite{liu2023verilogevalevaluatinglargelanguage,pinckney2025revisitingverilogevalyearimprovements}. The performance of this work rivals that of current LLM-based solutions for Verilog code generation. Although the benchmark provides an adequate analysis for \textit{pass@k} metrics, it does not cover multimodule designs; therefore, we created a small benchmark of complex designs to test with \emph{DeepV}, showcasing this work's application for complex, hierarchical designs in Section~\ref{subsec:case_studies}.

\subsection{Quantitative Analysis on VerilogEval} \label{subsec:verilog_eval_analysis}

\subsubsection{Experiment Settings} \label{ssubsec:exp_settings}
To quantitatively evaluate our framework, we conducted a large-scale experiment using the VerilogEval benchmark suite~\cite{liu2023verilogevalevaluatinglargelanguage,pinckney2025revisitingverilogevalyearimprovements}. VerilogEval is the standard benchmark for RTL code generation, comprising 156 human-generated problems of varying complexity that require the generation of a functionally correct RTL module from a natural language specification. To assess performance across different model architectures and sizes, a variety of back-end large language models were evaluated. These included open-source models, such as Mistral-7B-Instruct and CodeLlama-7B-Instruct, as well as a range of proprietary models from OpenAI, e.g., GPT-5 Chat. The retrieval component of our framework utilized the all-MiniLM-L6-v2 sentence transformers model for generating embeddings.

All code generation calls to the LLM were made with consistent inference parameters to ensure fair comparison across models and configurations. The temperature was set to 0.8 to encourage a degree of creativity while maintaining structural coherence, and top-p was set to 0.95 for nucleus sampling. A maximum of 1500 new tokens was allocated for each generation to accommodate complex modules. Two configurations were also evaluated to measure the direct impact of our RAG approach: a \emph{DeepV} configuration with the full retrieval pipeline active, and a Baseline (Zero-shot) configuration where the retrieval mechanism was disabled along with lessened prompt engineering.

The evaluation on the VerilogEval~\cite{liu2023verilogevalevaluatinglargelanguage,pinckney2025revisitingverilogevalyearimprovements} was performed in two sequential stages: \textbf{syntax check} and \textbf{functional verification}. During syntax validation, each module generated by a model was compiled using the \textit{Icarus Verilog} (i.e., \textit{iVerilog}) tool~\cite{iverilog}. Afterwards, the module was tagged as \textit{Compiled} or \textit{Not Compiled}, depending on the outcome of the compilation. In the next step, syntactically correct modules were simulated with Verilator using the benchmark-provided testbenches. Next, modules were labeled as "Passed" only if the simulated results had zero mismatches with the reference solutions defined by the benchmark.

We carried out the quantitative assessment through the widely-used \textit{pass@k}~\cite{openai2021codex} metric. The metric provides the probability of a code solution passing validation under k independent generations. The calculation is based on Eq.~\ref{equ:pass@k}. 

\begin{equation}\label{equ:pass@k}
    \text{pass@}k = \mathbb{E} \left( 1 - \frac{\binom{n-c}{k}}{\binom{n}{k}} \right)
\end{equation}
\vspace{0.2in}

where \textit{n} is the number of iterations that a certain code generation has gone through, and \textit{c} is the number of codes that passed validation. 

\subsubsection{Analysis of RAG Improvement on LLMs}
\label{ssubsec:ragAnalysis}

To gauge the performance of \emph{DeepV}, we applied to framework to multiple models, both open source and proprietary. We observed the performance gains of our framework, backed by \emph{VerilogDB}~\cite{calzada2025verilogdblargesthighestqualitydataset}, compared to the baseline models to which we applied \emph{DeepV} to. Table~\ref{table: Syntax Comparison} and Table~\ref{table: Functionality comparison} summarize our findings on syntax correctness and functional accuracy of the LLM-generated modules from VerilogEval~\cite{liu2023verilogevalevaluatinglargelanguage,pinckney2025revisitingverilogevalyearimprovements}. 

With a very high-quality knowledge base provided to \emph{DeepV} models, great improvement can be seen in the semantic understanding of Verilog code. We evaluate syntax accuracy for baseline commercial and open source LLMs and showcase the improvement seen with our approach. As can be seen in Table \ref{table: Syntax Comparison}, each LLM benefited from the context provided by additional documents with regard to syntax. GPT-5 Chat was augmented the most with an increased syntax accuracy on VerilogEval of almost 24\% with only one document retrieved.  Also, 100\% syntax accuracy for \textit{pass@5} was achieved with only one document retrieved. GPT-4o also augmented its accuracy on the benchmark from 74.1\% to 97.6\% with one document retrieved. Open source models saw the largest increases on three documents of around 8\% for CodeLlama 7B-Instruct and 11\% for Mistral 7B-Instruct from their baselines. 

\emph{DeepV's} ability to enhance LLM's generation of functionally accurate RTL code is evident from Table~\ref{table: Functionality comparison}. \emph{DeepV} improves the output of GPT-5 (Chat-Latest), OpenAI's latest model, by 16\% on \textit{pass@1}, and 13\% on \textit{pass@10}. Moreover, GPT-4o and  GPT-4.1 obtain  8.7\% and 8.9\% improvement on \textit{pass@1}, as well as 5.8\% and 5.1\% on \textit{pass@10}. Furthermore, Claude Sonnet 4 sees an increase of almost 10\% and 4\% on \textit{pass@1} and \textit{pass@10}. \emph{DeepV} also uplifts Gemini 2.5 Flash's results by 18.5\% and 7.1\% on \textit{pass@1} and \textit{pass@10}. Additionally, smaller open-source models such as Mistral-7B-Instruct had a 15.5\% improvement on \textit{pass@1} and CodeLlama-7B-Instact results increase by 9.6\% on \textit{pass@10}.

%%%% ____Syntaz table Start____

\begin{table}[htbp]
\centering
\caption{Syntax Correctness Comparison between Baseline Models and Corresponding \emph{DeepV} Approach on VerilogEval}
\label{table: Syntax Comparison}
\Huge                                    
\setlength{\tabcolsep}{3pt}               
\renewcommand{\arraystretch}{1.2}     

\begin{adjustbox}{max width=\textwidth}  
\begin{tabular}{|c|c|ccc|ccc|ccc|}
\hline
\multirow{2}{*}{\textbf{Model}} & \multirow{2}{*}{\textbf{Open Source}} & \multicolumn{3}{c|}{\textbf{Baseline}} & \multicolumn{3}{c|}{\textbf{DeepV}} & \multicolumn{3}{c|}{\textbf{Improvement}} \\ \cline{3-11} 
& & \textbf{pass@1} & \textbf{pass@5} & \textbf{pass@10} & \textbf{pass@1} & \textbf{pass@5} & \textbf{pass@10} & \textbf{pass@1} & \textbf{pass@5} & \textbf{pass@10} \\ \hline
Mistral-7B-Instruct & \cmark & 25.3 & 64.0 & 77.6 & 36.1 & 69.6 & 81.4 & 10.8 & 5.6 & 3.8 \\ 
CodeLlama-7B-Instruct & \cmark & 29.6 & 66.5 & 82.1 & 37.9 & 77.4 & 89.1 & 8.3 & 10.9 & 7.1 \\ 
% Qwen3-coder & \cmark & 91 & 97.8 & 98;7 & 96.8 & 99.1 & 99.4 & 5.77 & 1.27 & 0.64 \\ 
GPT-4o & \xmark & 74.1 & 76.7 & 77.6 & 97.9 & 98.7 & 98.7 & 23.8 & \cellcolor{bestimprov}\textbf{21.9} & \cellcolor{bestimprov}\textbf{21.2} \\ 
GPT-4.1 & \xmark & 74.4 & 86.7 & 89.7 & 98.1 & \cellcolor{bestperf}\textbf{100} & \cellcolor{bestperf}\textbf{100} & 23.7 & 13.3 & 10.3 \\ 
GPT-5 Chat & \xmark & 73.7 & 82.1 & 84.0 & \cellcolor{bestperf}\textbf{99.4} & \cellcolor{bestperf}\textbf{100} & \cellcolor{bestperf}\textbf{100} & \cellcolor{bestimprov}\textbf{25.6} & 17.9 & 16.0 \\ 
Claude Sonnet 4 & \xmark & 82.1 & 90.5 & 92.9 & \cellcolor{bestperf}\textbf{99.4} & \cellcolor{bestperf}\textbf{100} & \cellcolor{bestperf}\textbf{100} & 17.3 & 9.5 & 7.1 \\ 
Gemini 2.5 Flash & \xmark & 69.9 & 88.9 & 94.9 & 91.7 & 98.4 & 99.4 & 21.8  & 9.5 & 4.5 \\ 

\hline
\end{tabular}
\end{adjustbox}
\end{table}
%%%% ____Syntax table End____ %%%%

%%%____ Functional Comparison table (Start)____%%%%

\begin{table}[htbp]
\centering
\caption{Functional Correctness Comparison Between Baseline Models and Corresponding \emph{DeepV} Approach on VerilogEval}
\label{table: Functionality comparison}
\Huge                                    
\setlength{\tabcolsep}{3pt}               
\renewcommand{\arraystretch}{1.2}     

\begin{adjustbox}{max width=\textwidth} % only shrink if needed
\begin{tabular}{|c|c|ccc|ccc|ccc|}
\hline
\multirow{2}{*}{\textbf{Model}} & \multirow{2}{*}{\textbf{Open Source}} & \multicolumn{3}{c|}{\textbf{Baseline}} & \multicolumn{3}{c|}{\textbf{DeepV}} & \multicolumn{3}{c|}{\textbf{Improvement}} \\ \cline{3-11} 
& & \textbf{pass@1} & \textbf{pass@5} & \textbf{pass@10} & \textbf{pass@1} & \textbf{pass@5} & \textbf{pass@10} & \textbf{pass@1} & \textbf{pass@5} & \textbf{pass@10} \\ \hline
Mistral-7B-Instruct & \cmark & 7.1 & 12.3 & 14.7 & 22.6 & 25.7 & 26.3 & 15.5 & \cellcolor{bestimprov}\textbf{13.4} & 11.5 \\ 
CodeLlama-7B-Instruct & \cmark & 14.2 & 17.0 & 17.9 & 19.0 & 25.3 & 27.6 & 4.8 & 8.2 & 9.6 \\ 
% Qwen3-coder & \cmark & 64.7 & 73.5 & 76.9 & 64.7 & 74.3 & 77.6 & 0.0 & 0.8 & 0.6\\ 
GPT-4o & \xmark & 59.0 & 62.1 & 63.5 & 67.7 & 68.9 & 69.2 & 8.7 & 6.8 & 5.8 \\ 
GPT-4.1 & \xmark & 61.0 & 70.4 & 74.4 & 69.9 & 76.9 & 79.5 & 8.9 & 6.6 & 5.1 \\ 
GPT-5 Chat & \xmark & 60.9 & 68.8 & 69.9 & \cellcolor{bestperf}\textbf{76.9} & \cellcolor{bestperf}\textbf{81.3} & \cellcolor{bestperf}\textbf{83.3} & 16.0 & 12.4 & \cellcolor{bestimprov}\textbf{13.5} \\ 
Claude Sonnet 4 & \xmark & 66.0 & 75.5 & 78.2 & 75.6 & 80.6 & 82.1 & 9.6 & 5.1 & 3.8 \\ 
Gemini 2.5 Flash & \xmark & 53.2 & 69.2 & 73.7 & 71.8 & 78.4 & 80.8 & \cellcolor{bestimprov}\textbf{18.6} & 9.2 & 7.1 \\

\hline
\end{tabular}
\end{adjustbox}
\end{table}

%%%____ Functional Comparison table (Start)____%%%%

\subsubsection{Comparison with SOTA Techniques} \label{ssubsec:comparison}

Comparing the results of our framework to baseline models reveals a significant improvement in the performance of the LLM; however, it is also important to assess how effective our technique is in comparison to other state-of-the-art techniques. Table~\ref{tab:compare_soa_generators} depicts this comparison, where \emph{DeepV} outperforms the latest solutions in all metrics for the VerilogEval benchmark. \emph{DeepV} achieved a 76.9\% \textit{pass@1} value, outscaling the previously best scor, CraftRTL~\cite{craftrtl}, by 8.9\%. Moreover, \emph{DeepV} achieves an 81.3\% on \textit{pass@5}, surpassing Veriseek~\cite{wang2025largelanguagemodelverilog} by 4.4\%. Furthermore, our framework surpasses Veriseek~\cite{wang2025largelanguagemodelverilog} by 1.6\%, receiving an 83.3\% on \textit{pass@10}.

\begin{table}[htbp]
\centering
\caption{Performance Comparison of Functional Correctness on VerilogEval between Baseline Models, RTL Specific Solutions, and \emph{DeepV}. Results are reported for \textit{pass@1} at \texttt{temperature = 0.2}, and \textit{pass@5} and \textit{pass@10} at \texttt{temperature = 0.8}}
\label{tab:compare_soa_generators}
\renewcommand{\arraystretch}{1.2} 
\setlength{\tabcolsep}{6pt}      

\begin{tabular}{@{}|l|l|>{\centering\arraybackslash}p{1.8cm}|c|c|c|@{}} 
\toprule
\textbf{Category} & \textbf{Model / Method [Base Model]} & \textbf{Open Source} & \textbf{pass@1} & \textbf{pass@5} & \textbf{pass@10} \\
\midrule
\multirow{3}{*}{\textit{General Purpose}} & Mistral-7B-Instruct & \cmark & 7.1 & 12.3 & 14.7 \\
& CodeLLaMa-7B-Instruct & \cmark & 14.2 & 17.0 & 17.9 \\
% & Qwen3-Coder & \cmark & 64.7 & 73.5 & 76.9\\
\midrule
\multirow{3}{*}{\textit{Commercial}} & GPT-4o & \xmark & 59.0 & 62.1 & 63.5 \\
& GPT-4.1 & \xmark & 61.0 & 70.4 & 74.4 \\
& GPT-5 Chat & \xmark & 60.9 & 68.8 & 69.9 \\
\midrule
\multirow{12}{*}{\textit{RTL Specific}} & RTLCoder [DeepSeek-Coder-6.7B]~\cite{rtlcoder} & \cmark & 41.6 & 50.1 & 53.4 \\
& CodeV [Qwen-Coder]~\cite{codev} & \cmark & 59.2 & 65.8 & 69.1 \\
& Verigen [CodeLlama-7B]~\cite{verigen} & \cmark & 30.3 & 43.9 & 49.6 \\
& Origen [DeepSeek-Coder-7B]~\cite{origen} & \cmark & 54.4 & 60.1 & 64.2 \\
& RTL++ [CodeLlama-7B]~\cite{rtlpp} & \cmark & 59.9 & 68.8 & 72.1 \\
& BetterV [CodeQwen-7B]~\cite{betterv} & \xmark & 46.1 & 53.7 & 58.2 \\
& CraftRTL [StarCoder2-15B]~\cite{craftrtl} & \xmark & 68.0 & 72.4 & 74.6 \\
& VeriThoughts [Qwen-2.5-Instruct-14B]~\cite{yubeaton2025verithoughtsenablingautomatedverilog} & \cmark & 43.7 & 52.2 & 55.1 \\
& ReasoningV [DeepSeek-Coder-6.7B]~\cite{qin2025reasoningvefficientverilogcode} & \cmark & 57.8 & 69.3 & 72.4 \\
& Veriseek [DeepSeek-Coder-6.7B]~\cite{wang2025largelanguagemodelverilog} & \cmark & 61.6 & 76.9 & 81.7 \\
& ITERTL [DeepSeekV2-7k]~\cite{wu2025itertliterativeframeworkfinetuning} & \cmark & 53.8 & 60.8 & 64.1 \\
& AutoVCoder [CodeQwen-7B]~\cite{autovcoder} & \cmark & 48.5 & 55.9 & - \\

\midrule
% Corrected \multirow from 7 to 8 to match the number of models
\multirow{8}{*}{\textbf{DeepV}} & \textbf{DeepV[Mistral-7B-Instruct]}   & \xmark & \textbf{22.6} & \textbf{25.7} & \textbf{26.3} \\
& \textbf{DeepV[CodeLLaMa-7B-Instruct]} & \xmark & \textbf{19.0} & \textbf{25.3} & \textbf{27.6} \\
% & \textbf{DeepV[Qwen3-coder]}           & \xmark & \textbf{64.7} & \textbf{74.3} & \textbf{77.6} \\
& \textbf{DeepV[GPT-4o]}                & \xmark & \textbf{67.7} & \textbf{68.9} & \textbf{69.2} \\
& \textbf{DeepV[GPT-4.1]}               & \xmark & \textbf{69.9} & \textbf{76.9} & \textbf{79.5} \\
& \textbf{DeepV[GPT-5 Chat]}    & \xmark & \cellcolor{bestperf}\textbf{76.9} & \cellcolor{bestperf}\textbf{81.3} & \cellcolor{bestperf}\textbf{83.3} \\
& \textbf{DeepV[Claude Sonnet 4]}       & \xmark & \textbf{75.6} & \textbf{80.6} & \textbf{82.1} \\
& \textbf{DeepV[Gemini 2.5 flash]}      & \xmark & \textbf{71.8} & \textbf{78.4} & \textbf{80.8} \\

\bottomrule
\end{tabular}
\end{table}

\subsubsection{Ablation Study - RAG Configurations} \label{ssubsec:ablation_study}

\begin{figure*}[htbp]
    \vspace{0.2in}
    \centering
    % Row of 6 subfigures
    \begin{adjustbox}{max width=\textwidth}
        \begin{tabular}{cccc}
            \subcaptionbox{Mistral Syntax\label{fig:MistSyn}}{\includegraphics[width=0.25\textwidth]{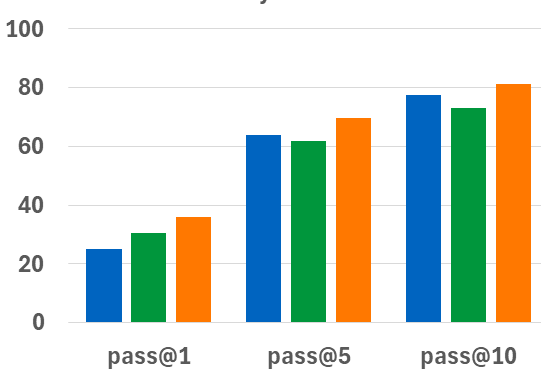}} &
            \subcaptionbox{Mistral Functionality\label{fig:MistFunc}}{\includegraphics[width=0.25\textwidth]{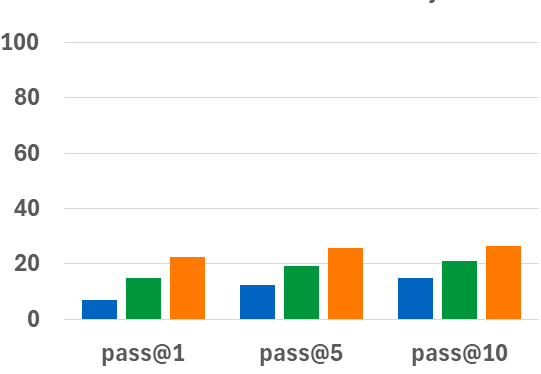}} &
            \subcaptionbox{CodeLLaMa Syntax\label{fig:CodeLSyn}}{\includegraphics[width=0.25\textwidth]{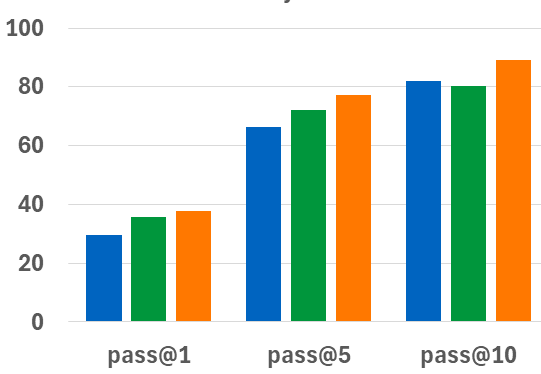}} &
            \subcaptionbox{CodeLLaMa Functionality\label{fig:CodeLFunc}}{\includegraphics[width=0.25\textwidth]{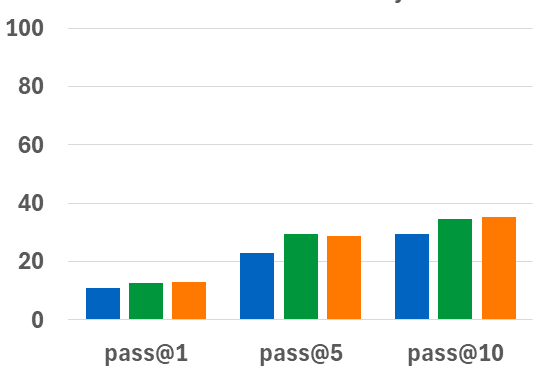}} \\
            &
            \subcaptionbox{GPT-4o Syntax\label{fig:GPT-4oSyn}}{\includegraphics[width=0.25\textwidth]{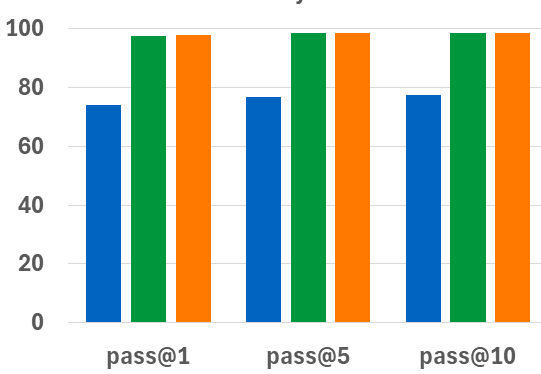}} &
            \subcaptionbox{GPT-4o Functionality\label{fig:GPT-4oFunc}}{\includegraphics[width=0.25\textwidth]{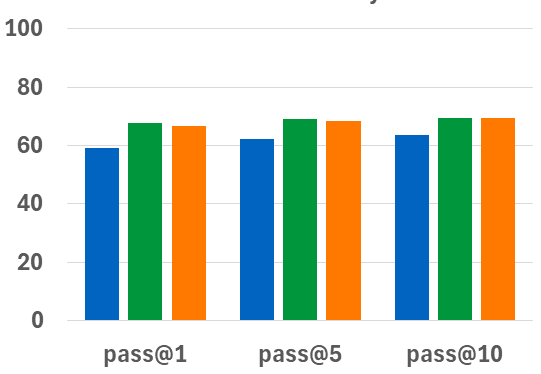}}
        \end{tabular}
    \end{adjustbox}

    % Legend below
    \vspace{0.2in}
    \includegraphics[width=0.3\textwidth]{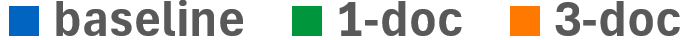}

    \caption{Comparison of Baseline Models with 1 Document and 3 Documents Retrieval by \emph{DeepV}. (Note: Mistral and CodeLLaMa models used are 7B Instruct)}
    \label{fig:1-doc,3-doc}
\end{figure*}

To assess the additional context provided by \emph{DeepV's} retrieved documents, we experimented with the number of documents retrieved by our RAG approach. During the initial phase, the understanding was that an increase in the number of retrieved documents would benefit LLMs by providing further context. This understanding was supported by the smaller open-source models' outputs. As shown in Fig.~\ref{fig:1-doc,3-doc}, both Mistral-7B-Instruct and CodeLLaMa-7B-Instruct gain approximately 9\% higher performance at the \textit{pass@10} metric for generating syntactically correct code, when three documents are retrieved over one document retrieval. Moreover, Mistral-7B-instruct's ability to produce functionally accurate documents improves by 8\% on \textit{pass@1}, as evident in Fig.~\ref{fig:MistFunc}. However, the GPT-4o results shown in Fig.~\ref{fig:GPT-4oSyn} and Fig.~\ref{fig:GPT-4oFunc} illustrate that the one-document retrieval performs almost equally to the three-document retrieval. 

This deviation from our initial findings compelled us to dig deeper into the retrieval system of the RAG. As a result, we applied dynamic sampling on the database with our RAG approach based on Algorithm~\ref{alg:dynamic_sampling}. We compared the dynamic retrieval system with only one-document retrieval. The results are depicted in figure~\ref{fig:1-doc,dynamic-sample}. The figures show that additional context over one-example retrieval does not necessarily benefit these models, but it helps for consistency to have a systematic way to organize the documents chosen to be augmented to each prompt, as opposed to having a hard-coded number of documents. As stated in Section~\ref{subsec:FAISS}, dynamic sampling has a dependency on the scoring system used for document and query vector comparison. Perhaps, only the first document provided enough context for most prompts, while a small margin of prompts required multiple documents; therefore, the 1 document context performed the same as the dynamic sampling. It can also be noted that LLMs typically have a maximum token limit for queries. In the case that the documents retrieved are lengthy codes, the full augmented query can be cut off. This can also apply to the previous ablation study results as seen in Fig.~\ref{fig:1-doc,3-doc}.

%%%%%______Dynamic Sampling Figures (Start)_____
\begin{figure*}[htbp]
    \centering
    % Row of 4 subfigures
    \begin{adjustbox}{max width=\textwidth}
        \begin{tabular}{cccc}
            \subcaptionbox{GPT-5 Chat Syntax\label{fig:GPT-5-Syn}}{\includegraphics[width=0.25\textwidth]{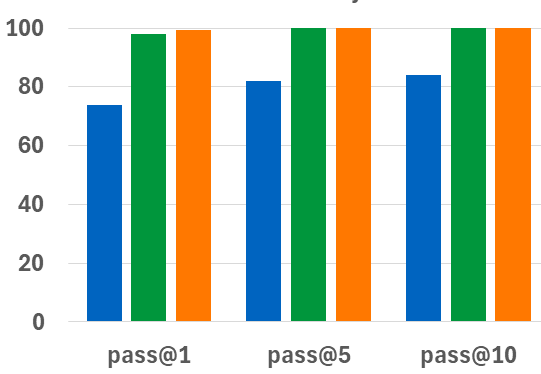}} &
            \subcaptionbox{GPT-5 Chat Functionality\label{fig:GPT-5-Func}}{\includegraphics[width=0.25\textwidth]{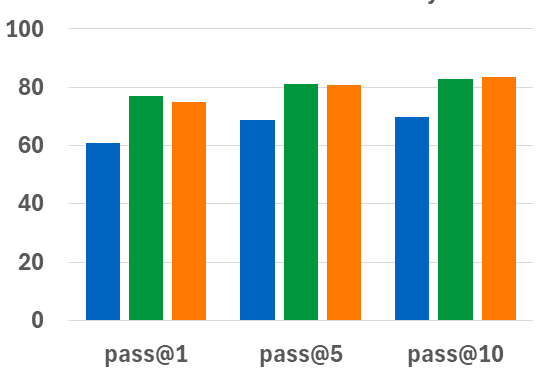}} &
            \subcaptionbox{GPT-4.1 Syntax\label{fig:GPT-4.1-Syn}}{\includegraphics[width=0.25\textwidth]{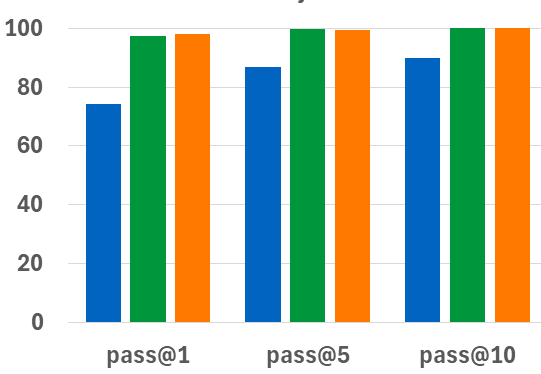}} &
            \subcaptionbox{GPT-4.1 Functionality\label{fig:GPT-4.1-Func}}{\includegraphics[width=0.25\textwidth]{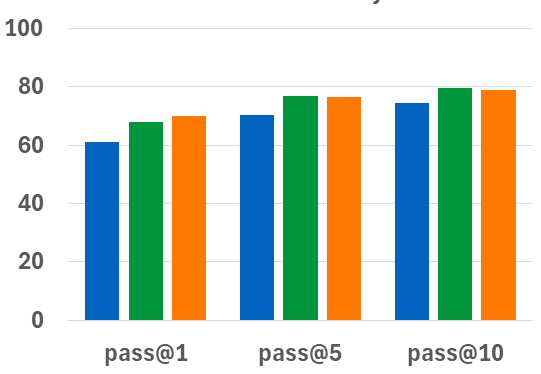}} \\
            \subcaptionbox{Claude Sonnet 4 Syntax\label{fig:Claude4-Syn}}{\includegraphics[width=0.25\textwidth]{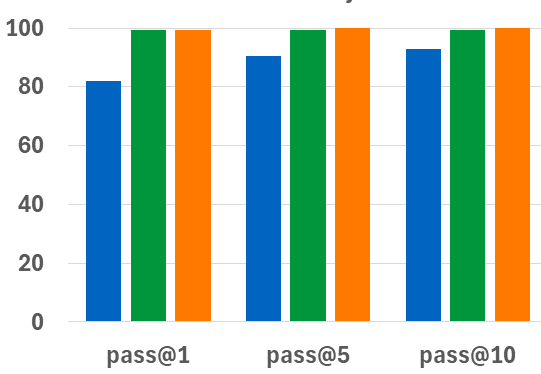}} &
            \subcaptionbox{Claude Sonnet 4 Functionality\label{fig:Claude4-Func}}{\includegraphics[width=0.25\textwidth]{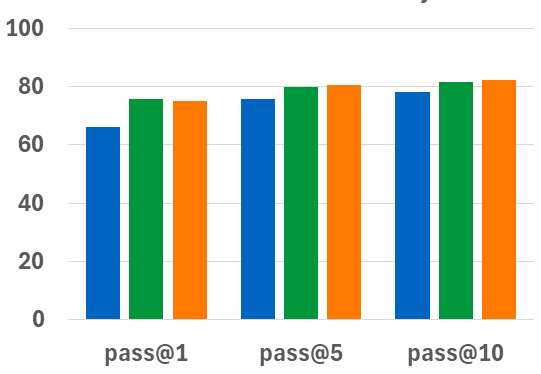}} &
            \subcaptionbox{Gemini 2.5 Flash Syntax\label{fig:Gemini-Syn}}{\includegraphics[width=0.25\textwidth]{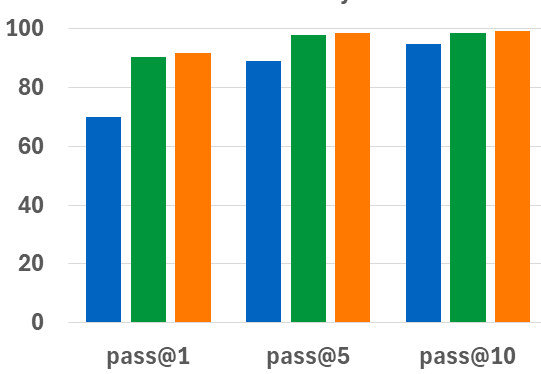}} &
            \subcaptionbox{Gemini 2.5 Flash Functionality\label{fig:Gemini-Func}}{\includegraphics[width=0.25\textwidth]{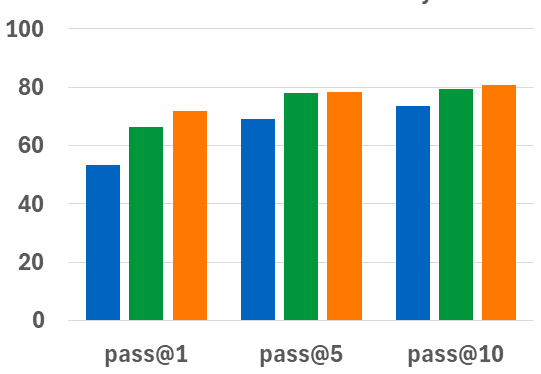}} 
        \end{tabular}
    \end{adjustbox}

    % Legend below
    \vspace{0.2in}
    \includegraphics[width=0.35\textwidth]{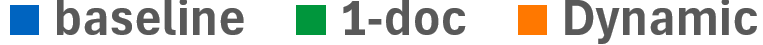}

    \caption{Comparison of LLM Performance between 1 Document Retrieval and Dynamic Sampling by \emph{DeepV}}
    \label{fig:1-doc,dynamic-sample}
    \vspace{0.2in}
\end{figure*}

%%%%%______Dynamic Sampling Figures (End)_____

\subsection{Case Studies: Complex Hierarchical IPs} \label{subsec:case_studies}

Despite the widely used benchmark, VerilogEval~\cite{liu2023verilogevalevaluatinglargelanguage, pinckney2025revisitingverilogevalyearimprovements}, having a variety of different design types and complexities, it does not have the prompts to test an LLM's capability to generate complex and hierarchical designs. To showcase that \emph{DeepV} can improve RTL generation of large hierarchical designs, we test a baseline GPT-5 Chat on four different hierarchical designs, which range in complexity but are multimodule. We then tested GPT-5 Chat with \emph{DeepV} for both 1-doc and dynamic sampling to assess the improvement on these specific complex cases. To configure GPT-5 Chat, a temperature of 1, top-p of 0.95, and max tokens of 10,000 were selected.

The four designs include a finite impulse response (FIR) filter used commonly in digital signal processing (DSP) applications to remove unwanted frequencies, a Sobel filter image processing algorithm used to detect edges, a secure random number generator (SRNG) used to create a cryptographically secure sequence of random bits, and a Universal Asynchronous Receiver/Transmitter (UART) module commonly used for system-level communication. The selected case study problems are listed in Table~\ref{tab:case_studies}, where it can be observed that the designs span different hierarchy depths, number of total modules, and application domains. We assess the application domains in order to test \emph{DeepV}'s capability to pull domain-specific information covering different use cases. The diversity in application domain is critical, because each domain possesses a unique set of constraints and implementation nuances; The knowledge required to correctly implement the timing and state logic of the UART protocol differs greatly from the algorithmic structure needed for a FIR filter or the entropic properties of a Secure RNG. By evaluating increased functional accuracy using \emph{DeepV} across all four varied designs, we highlight that \emph{DeepV}'s knowledge base is not narrowly specialized but is comprehensive and versatile. To summarize, these case studies demonstrate some of \emph{DeepV}'s key benefits:

\begin{enumerate}
    \item its ability to retrieve and apply correct, domain-specific context, proving its efficacy on a spectrum of real-world hardware engineering tasks.
    \item its ability to improve generation of complex, hierarchical IPs.
    \item its ability to adhere to formal specifications in prompts despite variations in context from the knowledge base.
\end{enumerate}

\begin{table*}[htbp]
\centering
\scriptsize
\caption{\centering{Designs selected for Case Study}}
\begin{adjustbox}{width=\linewidth,center}
\begin{tabular}{|c |c |c |c |c |c |c |c}
\hline \rule{0pt}{2ex} 
\textbf{ID} & \textbf{Design} & 
\textbf{Application Domain} &
\textbf{Hierarchy Depth} & \textbf{\# Modules} & \textbf{Generated Avg. Line Count} \\ [0.5ex] \hline \rule{0pt}{2ex}  
D1 & FIR Filter & Digital Signal Processing & Two & 10 & 202\\
\hline \rule{0pt}{2ex}     
D2 & Sobel Filter & Image Processing - Edge Detection & Two & 4 & 171\\
\hline \rule{0pt}{2ex} 
D3 & UART & Communication/Interface & Three & 5 & 285 \\
\hline \rule{0pt}{2ex} 
D4 & Secure RNG & Hardware Security & Four & 5 & 154 \\
\hline  
 
\end{tabular}
\end{adjustbox}
\label{tab:case_studies}
\end{table*}

\subsubsection{Case Study Setup and Execution}
% First, prompts for each design were prepared by identifying the port map and desired functionality. Each prompt was carefully designed to include the desired hierarchy that the LLM would design, including the depth and names of the different modules. Design considerations, such as timing, were taken into account as the testbench needs to enforce common tests for the design-under-test (DUT). Especially for UART, ensuring the prompt defines the precise interface timing specifications and frequencies is crucial. Testbenches were carefully designed based on the prompt information and intended functionality. Design samples were generated in 10 iterations per case study problem. Icarus-Verilog~\cite{iverilog} was used for compilation and simulation of the files.

To establish an unambiguous problem definition for the LLM, we first prepared detailed prompts for each design. Each prompt has a specified port map, core functionality, and the intended module hierarchy, including the names and depth of all sub-modules. This level of detail was crucial for guiding the LLM towards a structurally sound solution. Furthermore, design constraints were well-defined, e.g., for protocol-heavy modules like the UART, this included precise interface timing specifications and operating frequencies. To create a supporting validation framework, a functional testbench was developed for each prompt. This testbench served as the ground truth, enforcing the requirements laid out in the prompt and testing the design-under-test (DUT) for functional correctness. For each case study, we generated 10 design iterations to get a range of responses for the given prompt. Finally, each generated sample was compiled and simulated using \emph{iVerilog}~\cite{iverilog}, with a design deemed successful only if it compiled cleanly and passed all testbench assertions. Using this framework, we evaluated several configurations. For our baseline, we used GPT-5 Chat to generate 10 designs for each case study. We then applied \emph{DeepV} to generate an additional 10 designs per problem under two distinct retrieval settings: single-document (1-doc) and dynamic sampling. We define functional accuracy as the percentage of the 10 designs in a given set that successfully pass all simulation-based functional tests.

\begin{figure}[htbp]
    \centering
    \includegraphics[width=\textwidth]{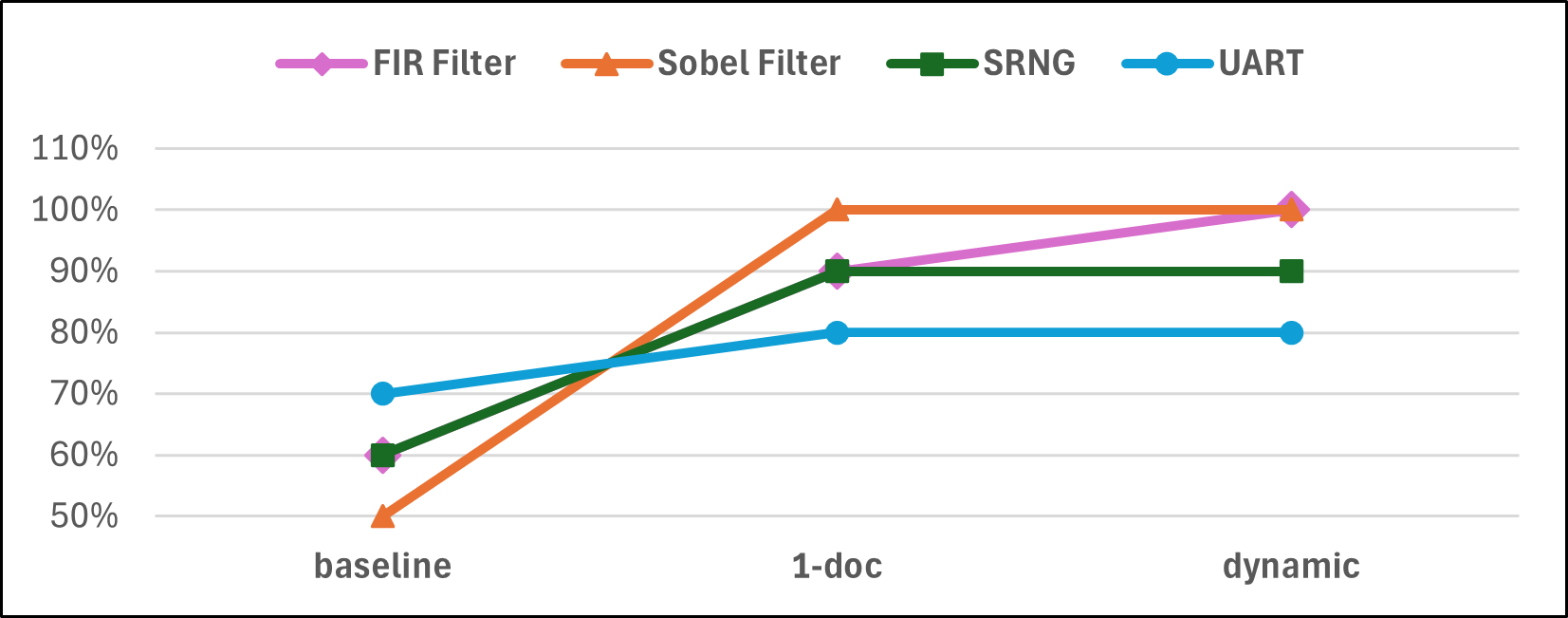}
    \caption{Case study functional accuracy across baseline (Gpt-5-chat-latest) and \emph{DeepV} configured for 1-doc and dynamic.}
    \label{fig:cs_plot}
\end{figure}

\subsubsection{Experiment Results}
% As mentioned, GPT-5 Chat generated 10 designs per case study problem as a baseline. With \emph{DeepV} applied, 10 designs were generated per case study problem for both 1-doc and dynamic retrieval settings. We calculate the functional accuracy as the number of generated codes that pass functional tests divided by 10 and multiplied by 100. 

% The Sobel filter problem saw the greatest increase in accuracy from the baseline to 1-doc of 50\%. Both the FIR filter and SRNG saw 30\% increases from the baseline to 1-doc, and accuracy was further increased on the FIR filter test case to 100\% using dynamic sampling. Testing on the UART test case saw the smallest gains, where a 10\% total enhancement was seen using \emph{DeepV}. As can be seen with the difference between the results of 1-doc and dynamic, three of the problems had functional accuracies which remained the same. This may be due to the dynamic RAG finding the first relevant document of an incredibly high similarity score compared to subsequent documents, so the same context was used between 1-doc and dynamic for these specific examples. The FIR filter example did prompt the dynamic RAG to find more similarly and highly scored documents which provided more context than the 1-doc alternative. As evident in the results, \textit{DeepV} can enhance LLMs' generation of functionally correct hierarchical designs, where even one provided document is sufficient to enhance functional accuracy.

The application of \emph{DeepV} resulted in significant improvements in functional correctness across all case studies. The most substantial gain was observed on the Sobel filter problem, which saw a 50\% accuracy increase with just 1-doc retrieval. Similarly, both the FIR filter and the SRNG experienced 30\% accuracy boosts from the baseline in the 1-doc setting. Notably, the FIR filter’s accuracy was further elevated to a perfect 100\% when using dynamic sampling. The UART test case, while showing the smallest gain, still saw a total enhancement of 10\%.

An interesting finding emerged when comparing the 1-doc and dynamic sampling results. For three of the four problems, the functional accuracy remained the same between the two settings. We hypothesize this is because the first retrieved document had a significantly higher similarity score than all subsequent candidates or other factors as mentioned in Section~\ref{ssubsec:ablation_study}, causing the dynamic sampling heuristic to be equivalent to that of the 1-doc results. In contrast, the FIR filter clearly benefited from the dynamic approach, as the retriever found multiple highly-scored documents that provided a richer context. Ultimately, these results demonstrate that \emph{DeepV} can substantially enhance an LLM's ability to generate functionally correct hierarchical designs, where even a single relevant document is often sufficient to see substantial accuracy gains.

\section{Discussion} \label{sec:discussion}

\subsection{Accessibility and Ease-of-Use}

\begin{figure}[htbp]
    \centering
    \includegraphics[width=\columnwidth]{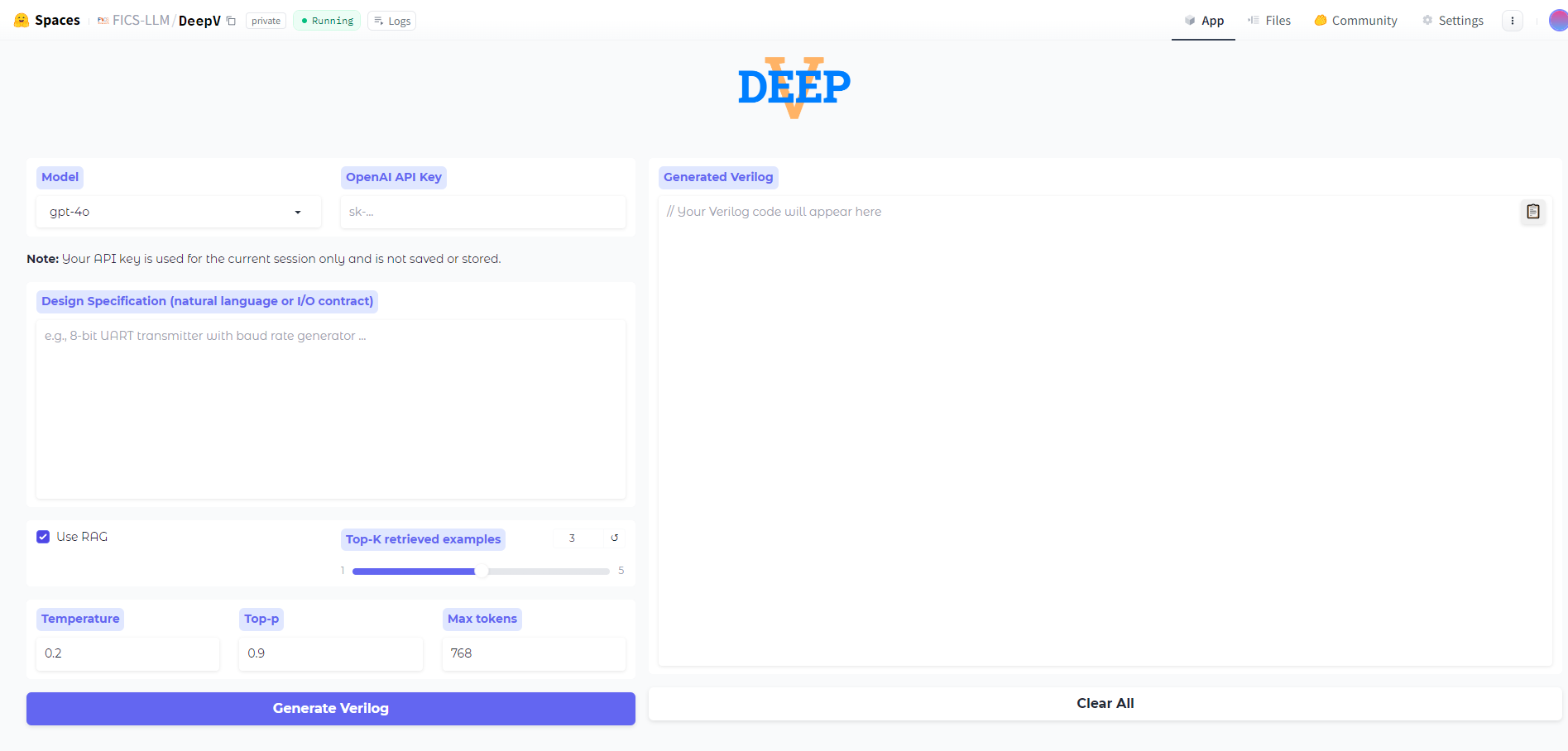}
    \caption{DeepV Hugging Face Space Application}
\label{fig:web_app}
\end{figure}

To allow the community to access \emph{DeepV}, we have integrated \emph{DeepV} into a Hugging Face (HF) Space: \url{https://huggingface.co/spaces/FICS-LLM/DeepV}.

Our \emph{DeepV} HF Space provides a web-based interface requiring no installations or specialized hardware. This removes barriers to experimentation and supports reproducible research. Firstly, the application allows the user to select an OpenAI model (GPT-4o, GPT-4.1, and GPT-5 Chat) and apply their OpenAI API key. We note the inputted API key is only active for the current session and is not saved or stored due to security concerns. Secondly, the user can input a design specification in an open format, unlike with fine-tuned LLMs, which each require specific input format requirements. Thirdly, the application allows for \emph{DeepV} configuration options like enabling RAG, number of retrieved documents, and LLM parameters, including temperature, top-p, and max tokens. As \emph{DeepV} is hosted on HF Spaces as a Gradio application, the Gradio client Python package can be run locally to interact with the application in an automated fashion. Due to this, \emph{DeepV} can be integrated into larger systems or agentic workflows. A tutorial API access Python script is provided in the \emph{DeepV} HF Space repository.

\subsection{Key Takeaways}
The results presented in this work offer several key takeaways for the field of LLM-aided hardware design.

\begin{keytakeaway}{$T_1$}
A high-quality knowledge base is as critical as model architecture; RAG can elevate general-purpose LLMs to rival specialized, fine-tuned models.
\end{keytakeaway}

Primarily, our findings demonstrate that a meticulously curated knowledge base combined with a sophisticated RAG framework can elevate the performance of general-purpose LLMs to a level competitive with, and in some cases superior to, highly specialized, fine-tuned models as seen in Table~\ref{tab:compare_soa_generators}. This suggests that for many RTL generation tasks, the bottleneck is not the model's inherent reasoning capability but its access to relevant, high-quality domain knowledge. \emph{DeepV}'s success, particularly with the latest GPT-5 Chat model, shows a potential path where continuous knowledge base refinement becomes as important as advancements in model architecture itself.

\begin{keytakeaway}{$T_2$}
Contextual needs are query-dependent; while a single example is often powerful, dynamic, query-aware retrieval is crucial for maximizing accuracy on complex designs.
\end{keytakeaway}

Furthermore, the comparison between 1-doc and dynamic sampling reveals nuances in how LLMs utilize context. While a single, highly relevant example provides a significant performance boost, dynamic sampling observed in several test cases suggests that more context is not always better context. The effectiveness of dynamic sampling on the FIR filter, however, proves that for certain complex problems, providing a wide range of context is critical for achieving high functional accuracy. This highlights the need for intelligent, query-aware retrieval strategies over fixed-context approaches, which adapt to the specific demands of each query.

\begin{keytakeaway}{$T_3$}
The effectiveness of RAG in hardware design is fundamentally tied to the quality of its knowledge base. Using syntactically correct, synthesizable code is highly important for generating reliable RTL.
\end{keytakeaway}

A central pillar of the \emph{DeepV} framework is its reliance on a meticulously curated knowledge base, which is a critical factor in its success. The framework utilizes \emph{VerilogDB}, a dataset composed of over 20,000 Verilog modules sourced from reliable repositories like GitHub and OpenCores, as well as academic materials~\cite{calzada2025verilogdblargesthighestqualitydataset}. Unlike RAG systems that may pull from unvetted sources, every module in \emph{VerilogDB} has undergone a rigorous preprocessing framework to confirm that it is both syntactically correct and synthesizable with standard EDA tools. This pre-verification makes it possible for the LLM to be grounded in industry-ready, error-free examples. Furthermore, the vectorization process creates a structured document for each module, combining the full Verilog code with crucial metadata such as natural language descriptions and detailed port lists, such that the retriever has rich semantic context for every search.

\begin{keytakeaway}{$T_4$}
A model-agnostic design, coupled with public deployment via a user-friendly interface, is important for driving adoption, reproducibility, and integration of new research into practical workflows.
\end{keytakeaway}

To maximize impact and utility, \emph{DeepV} was intentionally designed as a model-agnostic RAG framework that does not depend on a specific, fine-tuned model. This flexibility allows the framework to continuously benefit from the rapid advancements in general-purpose commercial LLMs without requiring costly retraining. To further promote adoption and reproducible research, \emph{DeepV} has been integrated into a publicly accessible HF Space, providing a web-based interface that requires no installations or specialized hardware from the user. This accessibility lowers the barrier to entry for experimentation and allows for seamless integration into larger systems or agentic workflows through its API. By providing an open and easy-to-use tool, we encourage community engagement and build a foundation for future RAG-centric research in hardware design.

\subsection{Future Work}
The success of the \emph{DeepV} framework in generating complex, hierarchical designs opens up several avenues for future research in LLM-driven EDA. One promising direction is the exploration of more sophisticated retrieval and scoring mechanisms. While the current system effectively uses semantic similarity with the \texttt{all-MiniLM-L6-v2} embeddings model, future iterations could incorporate hardware-aware metrics, such as graph-based similarity on dataflow representations or embeddings trained specifically on Verilog abstract syntax trees. This could further enhance the retriever's ability to identify the most functionally relevant examples for highly complex user queries as opposed to syntactical similarities.

Furthermore, \emph{DeepV}'s model-agnostic nature makes it an ideal component for integration into larger, agentic workflows. By equipping an autonomous agent with \emph{DeepV}'s high-quality retrieval capabilities, future systems could automate not only RTL generation but also subsequent tasks like self-correction, testbench generation, and synthesis optimization, creating a more holistic design automation solution. Finally, the core principles of \emph{DeepV} could be extended beyond code generation. The framework's ability to ground LLMs in a high-quality, verifiable knowledge base like \emph{VerilogDB}~\cite{calzada2025verilogdblargesthighestqualitydataset} could be adapted to address other critical challenges in the hardware lifecycle, such as secure linting, functional verification, and more.

\section{Conclusion} \label{sec:conclusion}
The desire to propel EDA further has led hardware engineers to pursue LLM-based techniques for automated generation of RTL code. However, LLMs have not been trained on an adequate amount of hardware knowledge to have the capability in producing syntax-free, functionally accurate codes based on a given query. Moreover, fine-tuning these LLMs on hardware knowledge has limitations with the constant advancements of baseline models. In this work, we show that with an extensive knowledge base, a RAG framework can improve the performance of commercial LLMs by a large margin. Our work, \emph{DeepV}, is model-agnostic, can be continuously updated, and provide high functional accuracy on a premier benchmark in this field, VerilogEval. As EDA automation moves forward with LLM-based frameworks, \emph{DeepV} can act as a valuable tool not just for the accurate creation of RTL code but also for continued research within this community with our available-to-use HF space.

\bibliographystyle{IEEEtran}
\bibliography{deepv.bib}
\end{document}